\documentclass[aps,prd,onecolumn,groupedaddress,showpacs,nofootinbib,amssymb]{revtex4}
\usepackage[dvips]{graphicx}
\usepackage{amssymb}
\usepackage{amsmath}
\usepackage{graphicx}
\usepackage{amsfonts}
\usepackage{bm}
\begin{document}

\title{New Traversable Wormholes Solutions in $f(T)$ Gravity }
\author{R.~C.~Tefo,$^{(a)}$\,\footnote{Email: teffcas@gmail.com}
P.~H.~Logbo,$^{(a,b)}$\,\footnote{Email: pascoloo@yahoo.fr}
M.~J.~S.~Houndjo,$^{(a,b)}$\,\footnote{Email: sthoundjo@yahoo.fr}
J.~Tossa,$^{(a)}$\,\footnote{Email: joel.tossa@imsp-uac.org}}
\affiliation{ 
$^{a}$ \, Institut de Math\'{e}matiques et de Sciences Physiques, 01 BP 613,  Porto-Novo, B\'{e}nin\\
$^{b}$\, Facult\'e des Sciences et Techniques de Natitingou, BP 72, Natitingou, B\'enin 
}
\begin{abstract}
In this paper we search for dynamical traversable wormhole solution in the modified $f(T)$ theory of gravity, $T$ being the torsion scalar. For such wormhole, the time dependence is inserted in the static  traversable wormhole metric of Morris and Thore. Two set of tetrads are adopted: the  diagonal and the non-diagonal tetrads. The diagonal set of tetrads constrains and reduces $f(T)$ model to teleparallel theory where usual solutions have been found. With diagonal set of tetrads, free from the teleparallel theory constraint, our results show that the existence of traversable wormhole is possible only for non-dynamical spacetime metric, i.e, static traversable wormhole solutions. Moreover we take into account energy condition analysis and the results show that the violation of NEC is not determinant for existence of static traversable wormhole solutions.   
\end{abstract}

%PACS numbers: 04.50.Kd, 95.36.+x, 98.80.-k, 98.80.Cq

\maketitle
%\makeatletter
%\renewcommand{\theequation}{\Roman{section}\,\Alph{subsection}.\arabic{equation}}
%\@addtoreset{equation}{subsection}{section}
%\makeatother

%\makeatletter
%\renewcommand{\theequation}{\Roman{section}.\arabic{equation}}
%\@addtoreset{equation}{section}
%\makeatother

\def\pp{{\, \mid \hskip -1.5mm =}}
\def\cL{\mathcal{L}}
\def\be{\begin{equation}}
\def\ee{\end{equation}}
\def\bea{\begin{eqnarray}}
\def\eea{\end{eqnarray}}
\def\tr{\mathrm{tr}\, }
\def\nn{\nonumber \\}
\def\e{\mathrm{e}}

%\tableofcontents

\section{Introduction}
Wormhole is known as a hypothetical object behaving like a tunnel or bridge that provides a connection between two different universes. 
The main challenge about the existence of wormhole is generally focused on the energy conditions and appears as  an interesting topic is gravitation. Traversable wormhole concept has first introduced by Morris and Thore \cite{18dezia}. It is then important to point out that this idea is quite different from the previous concept of Einstein-Rosen bridge \cite{19dezia} and also was unlike the concept of charge-carrying microscopic wormholes developed by Wheeler \cite{20dezia}. Various works have been performed in this way such, dynamical wormhole solutions \cite{8desharif}, brane wormholes \cite{10desharif}, traversable wormholes \cite{9desharif}, in the optic to minimize the violation of the null energy condition (NEC).  Specifically, the study of wormhole in the context of modified theories of gravity has received more attention. The wormhole geometries has been performed in $f(R)$ gravity in \cite{25dezia}, where some exact solutions according to captivated shape-function have been found. Still in $f(R)$ theory of gravity, static wormhole solution were explored where it is concluded that wormhole solution is possible in certain regions for only barotropic matter case \cite{26dezia}. In the context of $f(R,T)$ theory of gravity, static spherically wormhole solutions have been discussed in \cite{27dezia}. We characterize static wormhole by the metric  (\ref{chakrametric}) where $\Phi(r)$ is known as red-shift function essentially important for the determination of the event horizon \cite{chakra}. The function $B(r)$
is called the shape function and is representative of the spatial shape of the wormhole. It is well known in the literature that wormhole geometries is dictated by the red-shift and shape functions through the spacetime metric, while the field equations are used to evaluate the corresponding matter content. In general the matter content violates the NEC \cite{7dechakra,8dechakra,9dechakra,10dechakra,11dechakra,12dechakra,33dechakra} and this violation is essentially  due to its topology \cite{13dechakra,14dechakra}. The most interesting feature for the wormhole is the traversability characterized by the absence of horizon, meaning that the function $e^{2\phi(r)}$ is finite everywhere, and the condition $B(r_0)=r_0$ holds at the throat $r=r_0$. \par
Our goal in this paper is to search for relativistic wormholes (dynamical wormhole) in the framework of $f(T)$ theory of gravity. This kind of wormhole comes from the inclusion of the scale factor in the Morris-Thorne metric evidently known as static wormhole spacetime metric. For this kind of metric, and in general, including matter content, their existence satisfies weak energy condition (WEC) and dominate energy condition (DEC) \cite{20dechakra,21dechakra,22dechakra,23dechakra}. Several works based on traversable wormhole  have been performed in other king of modified theories of gravity; in $f(R)$ theory of gravity \cite{39dechakra,42dechakra}, in $f(G)$  theory of gravity \cite{34dechakra,35dechakra,36dechakra,37dechakra,38dechakra}, and interesting results have been found.
 For our purpose we consider the dynamical traversable wormhole metric and search for possible physical solutions where we make use for both diagonal and non-diagonal set of tetrads. For diagonal tetrads, $f(T)$ model is constrained to Teleparallel where there is no more interesting solution. Attention is therefore attached to non-diagonal set of tetrads. Our results show that the possible traversable wormhole solutions far from the Teleparallel theory, i.e, $f_{TT}(T)\neq 0$, are favored by absence of time dependence of the metric, meaning that only static traversable wormholes are possible in $f(T)$ theory of gravity.\par
The paper is organized as follows: the Sec.\ref{sec2} is devoted to the generality on $f(T)$ theory of gravity and the introduction to the wormhole metric, where we search for traversable wormhole solution within diagonal set of tetrads. In Sec.\ref{sec3}, we attach attention to a set of non-diagonal tetrads and search for traversable wormhole solutions. The analysis about energy conditions is performed in Sec.\ref{sec4}, and the conclusion in Sec.\ref{sec5}. 
%%%%%%%%%%%%%%%%%%%%%%%%%%%%%%%%%%%%%%%%%%%%%%%%%%%%%%%%%%%%%%%%%%%%%%%%%%%%%%%%%%%%%%%%%%%%%%%%%
\section{Generality on $f(T)$ gravity and wormhole metric}\label{sec2}
%%%%%%%%%%%%%%%%%%%%%%%%%%%%%%%%%%%%%%%%%%%%%%%%%%%%%%%%%%%%%%%%%%%%%%%%%%%%%%%%%%%%%%%%%%%%%%%%%%%%%
We start by the mathematical concept of $f(T)$ gravity based on the Weitzenbock's geometry. The Latin subscripts describe the elements of the tangent space to the manifold (spacetime), while the Greek ones are directly related to the spacetime. For a general spacetime metric, we express the line element as
\begin{equation}
ds^{2}=g_{\mu\nu}dx^{\mu}dx^{\nu}\; .\label{gmetric}
\end{equation} 
 The metric (\ref{gmetric}) can be projected onto the tangent space to the manifold such that the line element is written using the so-called tetrad matrix, as  
\begin{eqnarray}
ds^{2} &=&g_{\mu\nu}dx^{\mu}dx^{\nu}=\eta_{ij}\theta^{i}\theta^{j}\label{1}\; ,\\
dx^{\mu}& =&e_{i}^{\;\;\mu}\theta^{i}\; , \; \theta^{i}=e^{i}_{\;\;\mu}dx^{\mu}\label{2}\; ,
\end{eqnarray} 
where $\eta_{ij}=diag[1,-1,-1,-1]$ and $e_{i}^{\;\;\mu}e^{i}_{\;\;\nu}=\delta^{\mu}_{\nu}$ or  $e_{i}^{\;\;\mu}e^{j}_{\;\;\mu}=\delta^{j}_{i}$. This leads to $\sqrt{-g}=\det{\left[e^{i}_{\;\;\mu}\right]}=e$.  The spacetime can now be described by using  the tetrad matrix, where  Weitzenbock's connections are defined as 
\begin{eqnarray}
\Gamma^{\alpha}_{\mu\nu}=e_{i}^{\;\;\alpha}\partial_{\nu}e^{i}_{\;\;\mu}\label{wgamma}\; .
\end{eqnarray}
From the definition (\ref{wgamma}), it is obvious to note that the spacetime's curvature vanishes and only the torsion and its related quantities have contribution for this geometry. The antisymmetric property of the connection favors the torsion tensor, whose components are 
\begin{eqnarray}
T^{\alpha}_{\;\;\mu\nu}&=&\Gamma^{\alpha}_{\nu\mu}-\Gamma^{\alpha}_{\mu\nu}=e_{i}^{\;\;\alpha}\left(\partial_{\mu} e^{i}_{\;\;\nu}-\partial_{\nu} e^{i}_{\;\;\mu}\right)\label{torsion}\;.
\end{eqnarray}
The related tensors, the contorsion and the tensor $S$, are expressed through their components by  
\begin{eqnarray}
K^{\mu\nu}_{\;\;\;\;\alpha}&=&-\frac{1}{2}\left(T^{\mu\nu}_{\;\;\;\;\alpha}-T^{\nu\mu}_{\;\;\;\;\alpha}-T_{\alpha}^{\;\;\mu\nu}\right)\label{cont}\; ,\\
S_{\alpha}^{\;\;\mu\nu}&=&\frac{1}{2}\left( K_{\;\;\;\;\alpha}^{\mu\nu}+\delta^{\mu}_{\alpha}T^{\beta\nu}_{\;\;\;\;\beta}-\delta^{\nu}_{\alpha}T^{\beta\mu}_{\;\;\;\;\beta}\right)\label{stensor}\;.
\end{eqnarray}
The torsion scalar $T$ can now be expressed using (\ref{torsion})-(\ref{stensor}), as  
\begin{eqnarray}
T=T^{\alpha}_{\;\;\mu\nu}S^{\;\;\mu\nu}_{\alpha}\label{tore}\; .
\end{eqnarray}
We define the action of the $f(T)$ gravity as
\begin{eqnarray}
S[e^{i}_{\mu},\Phi_{A}]=\int\; d^{4}x\;e\left[\frac{1}{16\pi}f(T)+\mathcal{L}_{Matter}\left(\Phi_{A}\right)\right]\label{ftaction}\; ,
\end{eqnarray}
where units $G=c=1$ have been used and the $\Phi_{A}$ denote the matter fields. Considering the action (\ref{ftaction}) as a functional of the tetrads  $e^{i}_{\;\;\mu}$ and the fields $\Phi_{A}$, and vanishing its variation with respect to   $e^{i}_{\nu}$, i.e. the principle of minimum action, one gets the following equation of motion  \cite{barrow}
\begin{eqnarray}
S^{\;\;\nu\rho}_{\mu}\partial_{\rho}Tf_{TT}+\left[e^{-1}e^{i}_{\mu}\partial_{\rho}\left(ee^{\;\;\alpha}_{i}S^{\;\;\nu\rho}_{\alpha}\right)+T^{\alpha}_{\;\;\lambda\mu}S^{\;\;\nu\lambda}_{\alpha}\right]f_{T}+\frac{1}{4}\delta^{\nu}_{\mu}f=4\pi\mathcal{T}^{\nu}_{\mu}\label{em}\; ,
\end{eqnarray}
where $\mathcal{T}^{\nu}_{\mu}$ denotes the stress tensor,  $f_{T}=d f(T)/d T$ and $f_{TT}=d^{2} f(T)/dT^{2}$. \par
The general wormhole metric is assumed as \cite{chakra}
\begin{eqnarray}
ds^2=e^{2\Phi(r)}dt^2-\frac{dr^2}{1-B(r)/r}-r^2\left(d\theta^2+\sin^2{\theta}d\phi^2\right)\label{chakrametric}\;.
\end{eqnarray}
For the wormhole to be traversable and dynamic, the metric to be assumed in this paper is the flat Friedmann-Robertson-Walker (FRW), whose line element is
\begin{eqnarray}
ds^2=dt^2-A^2(t)\left[ \frac{dr^2}{1-B(r)/r}+r^2\left(d\theta^2+ \sin^2{\theta}d\phi^2\right)\right]\,.\label{le}
\end{eqnarray}
For an anisotropic fluid, the energy momentum tensor is given by the expression
\begin{eqnarray}
\mathcal{T}_{\mu}^{\,\nu}=\left(\rho+p_t\right)u_{\mu}u^{\nu}-p_t\delta^{\nu}_{\mu}+\left(p_r-p_t\right)v_{\mu}v^{\nu}\;,
\end{eqnarray}
where $u_\mu$ is the four-velocity, $v_\mu$ the unitary space-like vector in the radial direction, $\rho$ the energy density, $p_r$ the pressure in the direction of $u_\mu$ (radial pressure) and $p_t$ the pressure orthogonal to $v_\mu$ (tangential pressure). Since we are assuming an anisotropic spherically symmetric matter, one has $p_r \neq p_t$ , such that their equality corresponds to an isotropic fluid sphere.
\begin{eqnarray}
\frac{\partial \rho}{\partial t}+H\left(3\rho+p_r+2p_t\right)&=&0\,\label{eqconttemp},\\
\frac{\partial p_r}{\partial r}+\frac{2}{r}\left(p_r-p_t  \right)&=&0\label{eqcontrad}
\end{eqnarray}
According to the line element (\ref{le}), the field equations from (\ref{em}) read
\begin{eqnarray}
\frac{T'}{rA^2}\left(\frac{B}{r}-1\right)f_{TT}+\left[\frac{1}{2r^2A^2}\left(B'+\frac{B}{r}-1\right)+3\left(\frac{\dot{A}}{A}\right)^2\right]f_T+\frac{1}{4}f&=&4\pi \rho\,\label{dens}\\
T'\frac{\dot{A}}{A}f_{TT}+\left[\frac{1}{r^2A^2}\left(\frac{B}{r}-\frac{1}{2}\right)+\frac{\ddot{A}}{A}+2\left(\frac{\dot{A}}{A}\right)^2\right]f_T+\frac{1}{4}f&=&-4\pi p_r\,,\label{presr}\\
\frac{1}{A}\left[ \dot{A}\dot{T}+\frac{T'}{2rA}\left(\frac{B}{r}-1 \right)\right]f_{TT}+\left[\frac{1}{4r^2A^2}\left(B'+\frac{B}{r}-2\right)+\frac{\ddot{A}}{A}+2\left(\frac{\dot{A}}{A}\right)^2\right]f_T+\frac{1}{4}f&=&-4\pi p_t\,.\label{prest}\\
\frac{\dot A}{A}T'f_{TT}&=&0\\
\frac{\dot T}{rA^2}\left(1-\frac{B}{r}\right)f_{TT}&=&0\\
\frac{\cot{\theta}}{2r^2A^2}\dot{T}f_{TT}&=&0\\
\frac{\cot{\theta}}{2r^2A^2}T'f_{TT}&=&0\\
\end{eqnarray}
%the trace version of the equation reads
%\begin{eqnarray}
%\left[\frac{1}{r^2A^2}\left(B'+\frac{2B}{r}-2\right)+3\frac{\ddot{A}}{A}+9\left(\frac{\dot{A}}{A}\right)^2\right]f_T+f&=& 4\pi\left(\rho-2p_t-p_r\right)\;.
%\end{eqnarray}

The torsion scalar is given by
\begin{eqnarray}
T(t,r)=-6\left(\frac{\dot{A}}{A}\right)^2-\frac{2}{r^2A^2}\left(\frac{B}{r}-1\right)\,.\label{torsionrt}
\end{eqnarray}
For interesting analysis, various cases can be observed:\par
%%%%%%%%%%%%%%%%%%%%%%%%%%%%%%%%%%%%%%%%%%%%%%%%%%%%%%%%%%%%%%%%%%%%%
\subsection{ First case: $f_{TT}=0$} 
%%%%%%%%%%%%%%%%%%%%%%%%%%%%%%%%%%%%%%%%%%%%%%%%%%%%%%%%%%%%%%%%%%%%
This condition means that the algebraic function $f(T)$ behaves as 
\begin{eqnarray}
f(T)= k_1T+k_2\;,
\end{eqnarray}
where $k_1$ and $k_2$ are constants, which is the tele-parallel gravitational term. Without loss of generality, let us fix $k_1=1$ and $k_2=0$, such that the field equations become
\begin{eqnarray}
{3}\left( \frac{\dot A}{A} \right)^2+\frac{B'}{r^2A^2}=8\pi \rho  \label{eqT0}\\
2\frac{\ddot A}{A}+\left(\frac{\dot A}{A}\right)^2+\frac{B}{r^3A^2}=-8\pi p_r \label{eqTr}\\
2\frac{\ddot A}{A}+\left(\frac{\dot A}{A}\right)^2-\frac{B}{2r^3A^2}+\frac{B'}{2r^2A^2}=-8\pi p_t\label{eqTt}
\end{eqnarray}
Here the equations of continuity related to the time (\ref{eqconttemp})-(\ref{eqcontrad}) are identically satisfied.

\subsubsection{Vacuum solutions}
In vacuum case, the energy density and the pressure vanish. Then, by using any one of the equations (\ref{eqT0})-(\ref{eqTt}), one gets
\begin{eqnarray}
B(r)=-kr^3\;,\quad \quad A(t)=\sqrt{k}t+k_0\,,
\end{eqnarray}
where $k>0$ and $k_0$ are integration constants. Therefore the line element (\ref{le}) becomes
\begin{eqnarray}
ds^2=dt^2-(kt^2)\left[ \frac{dr^2}{1+kr^2}+r^2\left(d\theta^2+ \sin^2{\theta}d\phi^2\right)\right]\,,\label{le1}
\end{eqnarray}
which is the FRW type solution. Then,  for the specific choice $B(r)=-kr^3$ of the shape function, with scale factor $A(t)=\sqrt{k}t+k_0$, dynamic traversable wormhole vacuum solution in teleparallel gravity is of type FRW line element.

\subsubsection{The case $B(r)=B_0r^3+r$\;,\quad $A(t)=A_0t^\alpha$}
The goal here is to write down the different expressions of the energy density, the pressure and the corresponding line element. Hence, using the adopted expressions of $B(r)$ and $A(t)$, one gets
\begin{eqnarray}
\rho(t,r)&=& \frac{1}{8\pi}\left[   \frac{3\alpha^2}{t^2}+\frac{1+3B_0r^2}{A_0^2r^2t^{2\alpha}} \right]\;,\\
p_r(t,r)&=& \frac{1}{8\pi}\left[ \frac{\alpha(2-3\alpha)}{t^2}-\frac{1+B_0r^2}{A_0^2r^2t^{2\alpha}}\right]\;,\\
p_t(t)&=& \frac{1}{8\pi}\left[ \frac{\alpha(2-3\alpha)}{t^2}-\frac{B_0}{A_0^2t^{2\alpha}}\right]
\end{eqnarray} 
The torsion scalar (\ref{torsionrt}) becomes
\begin{eqnarray}
T(t)=-\frac{6\alpha^2}{t^2}-\frac{2B_0}{A_0^2t^{2\alpha}}\;.
\end{eqnarray}
Note that for the parameter $\alpha=1$ the torsion scalar becomes constant. The line element (\ref{le}) becomes
\begin{eqnarray}
ds^2=dt^2-A_0^2t^{2\alpha}\left[ -\frac{dr^2}{B_0r^2}+r^2\left(d\theta^2+ \sin^2{\theta}d\phi^2\right)\right]\,,\label{le2}
\end{eqnarray}
Here it is straightforward to note that $B_0$ has to be fixed as a negative constant. We conclude here that in the teleprallel theory, it also appears that dynamic traversable wormhole non-vacuum solutions are FRW type solution, where the energy density and the radial pressure, are radial coordinate and cosmic time dependent, while the tangential pressure is only time dependent. 
%%%%%%%%%%%%%%%%%%%%%%%%%%%%%%%%%%%%%%%%%%%%%%%%%%%%%%%%%%%%%%%%%%%%%%%%%%%ùùù
\subsection{Second case:   $\dot{T}=0$ and \;\; $T'=0$}
%%%%%%%%%%%%%%%%%%%%%%%%%%%%%%%%%%%%%%%%%%%%%%%%%%%%%%%%%%%%%%%%%%%%%%%
In this case, the field equations read
\begin{eqnarray}
\left[\frac{1}{2r^2A^2}\left(B'+\frac{B}{r}-1\right)+3\left(\frac{\dot{A}}{A}\right)^2\right]f_T+\frac{1}{4}f&=&4\pi \rho\,\label{dens1}\\
\left[\frac{1}{r^2A^2}\left(\frac{B}{r}-\frac{1}{2}\right)+\frac{\ddot{A}}{A}+2\left(\frac{\dot{A}}{A}\right)^2\right]f_T+\frac{1}{4}f&=&-4\pi p_r\,,\label{presr1}\\
\left[\frac{1}{4r^2A^2}\left(B'+\frac{B}{r}-2\right)+\frac{\ddot{A}}{A}+2\left(\frac{\dot{A}}{A}\right)^2\right]f_T+\frac{1}{4}f&=&-4\pi p_t\,.\label{prest1}\;.
\end{eqnarray}
For the torsion scalar (\ref{torsionrt}) to be free from $t$ and $r$, it is straightforward to see that the couple $(A(t),B(r))=(A_0,B_0r^3+r)$ is a potential candidate. Since the torsion scalar is a constant, $T_0$, one gets $f_T=0$, such that the energy density, the radial and tangential pressures are constants. Setting $f(T_0)/16\pi=K$, one gets
\begin{eqnarray}
\rho= -p_r=-p_t=K
\end{eqnarray}
The corresponding line element can be obtained from (\ref{le2}) by setting $\alpha=0$.

\section{Wormhole solutions from non-diagonal tetrad}\label{sec3}

It appears clearly that the diagonal tetrad constrains the $f(T)$ model to the tele-parallel one and there is no new solution\footnote{The solutions are the same that should be obtained in tele-parallel theory of gravity.}. Then, one can search for a non-diagonal tetrad as follows:

\begin{eqnarray}\label{nontet}
\{e^{a}_{\;\;\mu}\}=\left[\begin{array}{cccc}
1&0&0&0\\
0&\frac{A(t)}{\sqrt{1-\frac{B(r)}{r}}}\sin\theta\cos\phi &A(t)r\cos\theta\cos\phi &-A(t)r\sin\theta\sin\phi\\
0&\frac{A(t)}{\sqrt{1-\frac{B(r)}{r}}}\sin\theta\sin\phi &A(t)r\cos\theta\sin\phi &A(t)r\sin\theta\cos\phi \\
0&\frac{A(t)}{\sqrt{1-\frac{B(r)}{r}}}\cos\theta &-A(t)r\sin\theta &0
\end{array}\right]\;,
\end{eqnarray}  
whose inverse is 
\begin{eqnarray}\label{invnontet}
\{e_{a}^{\;\;\mu}\}=\left[\begin{array}{cccc}
1&0&0&0\\
0&[A(t)]^{-1}\sqrt{1-\frac{B(r)}{r}}\sin\theta\cos\phi &[A(t)]^{-1}\sqrt{1-\frac{B(r)}{r}}\sin\theta\sin\phi &[A(t)]^{-1}\sqrt{1-\frac{B(r)}{r}}\cos\theta\\
0&[rA(t)]^{-1}\cos\theta\cos\phi &[rA(t)]^{-1}\cos\theta\sin\phi &-[rA(t)]^{-1}\sin\theta \\
0&-[rA(t)]^{-1}\sin^{-1}\theta\sin\phi &[rA(t)]^{-1}\sin^{-1}\theta\cos\phi &0
\end{array}\right]\;.
\end{eqnarray}
Therefore, the field equations read
\begin{eqnarray}
\frac{T'}{rA^2}\left(-\frac{B}{r\sqrt{1-\frac{B}{r}}}+\frac{1}{\sqrt{1-\frac{B}{r}}}+\frac{B}{r}-1\right)f_{TT}+\nonumber\\
+\frac{1}{A^2}\left[\frac{1}{r^2}\left(\frac{B'}{2}+\frac{B}{2r}+\sqrt{1-\frac{B}{r}}-1\right)+3\dot{A}^2\right]f_T+\frac{f}{4}&=&4\pi \rho\,,\label{ndeq1}\\
\frac{\dot{A}}{A}\dot{T}f_{TT}+\frac{1}{A^2}\left[\frac{1}{r^2}\left(\frac{B}{r}+\sqrt{1-\frac{B}{r}}-1\right)+A\ddot{A}+2\dot{A}^2   \right]f_T+\frac{f}{4}&=&-4\pi p_r\,,\label{ndeq2}\\
\left[\frac{T'}{rA^2}\left(-\frac{B}{2r\sqrt{1-\frac{B}{r}}}+\frac{1}{2\sqrt{1-\frac{B}{r}}}+\frac{B}{2r}-\frac{1}{2}\right)+
\frac{\dot{A}}{A}\dot{T}\right]f_{TT}+\nonumber\\
\frac{1}{A^2}\left[\frac{1}{r^2}\left(\frac{B'}{4}+\frac{B}{4r}+\sqrt{1-\frac{B}{r}}-1\right)+A\ddot{A}+2\dot{A}^2   \right]f_T+\frac{f}{4}&=&-4\pi p_t\;,\label{ndeq3}\\
\frac{\dot A}{A}T'f_{TT}&=&0\;,\label{ndeq4}\\
\frac{\dot T}{rA^2}\left(\frac{B}{r\sqrt{1-\frac{B}{r}}}-\frac{1}{\sqrt{1-\frac{B}{r}}}-\frac{B}{r}+1\right)f_{TT}&=&0\label{ndeq5}
\end{eqnarray}
Here the torsion scalar reads
\begin{eqnarray}
T(t,r)= -6\left(\frac{\dot A}{A}\right)^2+\frac{2}{r^2A^2}\left(2-\frac{B}{r}-2\sqrt{1-\frac{B}{r}}\right)\;.\label{ndtrace}
\end{eqnarray}
From the equation (\ref{ndeq4}) it appears that by fixing $A(t)$ as a constant, i.e,  $A(t)=A_0$, the torsion scalar (\ref{ndtrace}) is free from the cosmic time and equations (\ref{ndeq4})-(\ref{ndeq5}) are satisfied without the need of the strong constraint $f(T)_{TT}=0$. Then the equations (\ref{ndeq1})-(\ref{ndeq3}) and the torsion scalar become,respectively
\begin{eqnarray}
\frac{T'}{A_0^2r}\left(-\frac{B}{r\sqrt{1-\frac{B}{r}}}+\frac{1}{\sqrt{1-\frac{B}{r}}}+\frac{B}{r}-1\right)f_{TT}+\nonumber\\
+\frac{1}{A_0^2r^2}\left(\frac{B'}{2}+\frac{B}{2r}+\sqrt{1-\frac{B}{r}}-1\right)f_T+\frac{f}{4}&=&4\pi \rho\,,\label{ndeq1r}\\
\frac{1}{A_0^2r^2}\left(\frac{B}{r}+\sqrt{1-\frac{B}{r}}-1\right)f_T+\frac{f}{4}&=&-4\pi p_r\,,\label{ndeq2r}\\
\frac{T'}{A_0^2r}\left(-\frac{B}{2r\sqrt{1-\frac{B}{r}}}+\frac{1}{2\sqrt{1-\frac{B}{r}}}+\frac{B}{2r}-\frac{1}{2}\right)f_{TT}+\nonumber\\
\frac{1}{A_0^2r^2}\left(\frac{B'}{4}+\frac{B}{4r}+\sqrt{1-\frac{B}{r}}-1\right)f_T+\frac{f}{4}&=&-4\pi p_t\;,\label{ndeq3r}
\end{eqnarray}
and
\begin{eqnarray}
T(r)=\frac{2}{r^2A_0^2}\left(2-\frac{B}{r}-2\sqrt{1-\frac{B}{r}}\right)\;.\label{ndtracer}
\end{eqnarray}
As performed in the previous section, where the diagonal tetrads are used, one can also distinguish some cases.
\subsubsection{Null radial pressure case}
In this case, we refer to the equation (\ref{ndeq2r}) which becomes
\begin{eqnarray}
\frac{1}{A_0^2r^2}\left(\frac{B}{r}+\sqrt{1-\frac{B}{r}}-1\right)f_T+\frac{f}{4}=0\,,\label{tbsolve1}.
\end{eqnarray}
The fundamental task to be performed in this subsection and the coming ones is to choose an appropriate expression to the $f(T)$ model and find out the corresponding expression of the radial function $B(r)$; or having this latter and reconstruct the associated $f(T)$ model.\par
\vspace{0.25cm}
\begin{center}
$\bullet$ {\bf Fixing $f(T)$ as exponential function of the torsion scalar}\par
\end{center}
\vspace{0.25cm}
Let us assume an exponential $f(T)$ and search for the analytical solution for the radial function $B(r)$; that is $f(T)=e^{aT}$ where $a$ is a constant. Then, the solutions of (\ref{tbsolve1}) read
\begin{eqnarray}
B_{1,2}(r)=1-\frac{r}{2}-\frac{aA_0^2r^2}{4}\pm\frac{1}{4}\sqrt{\left(aA_0^2r^2+2r-4\right)^2-aA_0^2r^4\left(aA_0^2r^2-8\right)}
\label{solB1}
\end{eqnarray}
Therefore the respective expressions of the torsion scalar, the energy density and the tangential pressure by injecting (\ref{solB1}) into (\ref{ndtracer}), (\ref{ndeq1r}) and (\ref{ndeq3r}).  Due to their long expressions we let them for numerical analysis in future analysis.\par

\vspace{0.25cm}
\begin{center}
$\bullet$ {\bf Special case : $B(r)=-B_0^2r3+r$}\par
\end{center}
\vspace{0.25cm}
In this case, within the expression of $B(r)$, we refer to the equation  (\ref{ndtracer}) and (\ref{tbsolve1}), and try to reconstruct the corresponding $f(T)$ model. First, the torsion scalar (\ref{ndtracer}) becomes
\begin{eqnarray}
T(r)=\frac{2}{A_0^2}\left(B_0-\frac{1}{r}\right)^2\;\label{ndtracer1} 
\end{eqnarray}
and the corresponding equation of field related to the null radial pressure (\ref{tbsolve1}) reads
\begin{eqnarray}
-4\frac{B_0}{A_0}\sqrt{\frac{T}{2}}f_{T}+f=0\;,
\end{eqnarray}
whose general solution is
\begin{eqnarray}
f(T)=C_1e^{\frac{A_0}{B_0}\sqrt{\frac{T}{2}}}\,,\label{model1}
\end{eqnarray}
where $C_1$ is an integration constant. Thus, using the definition of $B(r)$ with (\ref{model1}), the expressions of energy density and the tangential pressure can be determined in terms of the radial coordinate $r$. These expressions will be used for the analysis of energy conditions in the section (\ref{sec4}). \par 
The solution obtained here should be generalized to the case of dust dominated universe, where, due to the vanishing radial and tangentail pressures, the energy conditions are naturally satisfied.

\subsubsection{General isotropic traversable Wormhole solutions}

Here we adopt $p_r=p_t$, still with the shape function $B(r)=-B_0^2r^3+r$. The resulted differential equation from (\ref{ndeq2r})-(\ref{ndeq3r}), without fixing $B(r)$, reads
\begin{eqnarray}
T'\left(-\frac{B}{2r\sqrt{1-\frac{B}{r}}}+\frac{1}{2\sqrt{1-\frac{B}{r}}}+\frac{B}{2r}-\frac{1}{2}\right)f_{TT}+
\frac{1}{r}\left(\frac{B'}{4}-\frac{3B}{4r}\right)f_T=0\;,
\end{eqnarray}
where, making use of the torsion scalar (\ref{ndtracer1}), one gets two independent differential equations as the choice $\sqrt{B^2r^2}=B_0r$ or $\sqrt{B^2r^2}=-B_0r$ is made. Thus, with the first choice, one obtains the following differential
\begin{eqnarray}
Tf_{TT}+\frac{1}{2}\left(1-\frac{A_0}{B_0}\sqrt{\frac{T}{2}}\right)f_T=0
\end{eqnarray}
whose general solution reads
\begin{eqnarray}
f(T)=\frac{2\sqrt{2}k_1B_0}{A_0}e^{\frac{A_0}{B_0}\sqrt{\frac{T}{2}}}+k_2\;,\label{gsol1}
\end{eqnarray}
where $k_1$ and $k_2$ are integration constants.\par 
With the second choice, one gets 
\begin{eqnarray}
4B_0\sqrt{\frac{T}{2}}\left(2B_0+A_0\sqrt{\frac{T}{2}}\right)f_{TT}-A_0\left(B_0+A_0\sqrt{\frac{T}{2}}\right)f_T=0\;,
\end{eqnarray}
whose general solution reads
\begin{eqnarray}
f(T)=\frac{2\sqrt{2}k_3B_0}{A_0^2}e^{\frac{A_0}{B_0}\sqrt{\frac{T}{2}}}-\frac{4\sqrt{2}k_4B_0}{A_0^2e^2}\mbox{ExpIntegralEi}{\left(2+\frac{A_0}{B_0}\sqrt{\frac{T}{2}}\right)}\;,\label{gsol2}
\end{eqnarray}
where $k_3$ and $k_4$ are integration constants, and this solution generalizes the previous one. Thus, by setting $k_3=A_0k_1$ and $k_4=k_2=0$, the solution (\ref{gsol1}) is recovered.

\section{Energy conditions applied to the solutions}\label{sec4}
The wormhole physic is essentially governed by  the violation of null energy condition. To this end, we can recast the field equations as follows
\begin{eqnarray}
\frac{B'}{2A_0^2r^2}&=&4\pi\rho_{eff}\;,\\
\frac{B}{2A_0^2r^3}&=&-4\pi p_{r,\;eff}\;,\\
-\frac{B}{2A_0^2r^3}+\frac{B'}{4A_0^2r^2}&=&-4\pi p_{t,\;eff}
\end{eqnarray}
with 
\begin{eqnarray}
\rho_{eff}&=&\frac{1}{4\pi f_T}\Bigg[ 4\pi\rho-\frac{T'}{A_0^2r}\left(-\frac{B}{r\sqrt{1-\frac{B}{r}}}+\frac{1}{\sqrt{1-\frac{B}{r}}}+\frac{B}{r}-1\right)f_{TT}
-\frac{1}{A_0^2r^2}\left(\frac{B}{2r}+\sqrt{1-\frac{B}{r}}-1\right)f_T-\frac{f}{4}\Bigg],\\
p_{r\;eff}&=&\frac{1}{4\pi f_T}\left[4\pi p_r+ \frac{1}{A_0^2r^2}\left(\frac{B}{2r}+\sqrt{1-\frac{B}{r}}-1\right)f_T+\frac{f}{4}\right]\;,\\
p_{t\;eff}&=&\frac{1}{4\pi f_T}\Bigg[4\pi p_t+\frac{T'}{A_0^2r}\left(-\frac{B}{2r\sqrt{1-\frac{B}{r}}}+\frac{1}{2\sqrt{1-\frac{B}{r}}}+\frac{B}{2r}-\frac{1}{2}\right)f_{TT}+\nonumber\\
&+&\frac{1}{A_0^2r^2}\left(\frac{B'}{4}+\frac{B}{2r}+\sqrt{1-\frac{B}{r}}-1\right)f_T+\frac{f}{4}\Bigg]
\end{eqnarray}
%%%%%%%%%%%%%%%%%%%%%%%%%%%%%%%%%%%%%%%%%%%%%%%%%%%%%%%%%%%%%%%%%%%%%%%%%%%%%%%%%%%%%%%%%%%%%%%%%%%%%%%%%%%%
%%%%%%%%%%%%%%%%%%%%%%%%%%%%%%%%%%%%%%%%%%%%%%%%%%%%%%%%%%%%%%%%%%%%%%%%%%%%%%%%%%%%%%%%%%%%%%%%%%%%%%%%%%%%
%%%%%%%%%%%%%%%%%%%%%%%%%%%%%%%%%%%%%%%%%%%%%%%%%%%%%%%%%%%%%%%%%%%%%%%%%%%%%%%%%%%%%%%%%%%%%%%%%%%%%%%%%%%%%%%
%%%%%%%%%%%%%%%%%%%%%%%%%%%%%%%%%%%%%%%%%%%%%%%%%%%%%%%%%%%%%%%%%%%%%%%%%%%%%%%%%%%%%%%%%%%%%%%%%%%%%%%%%%%%%%%
%%%%%%%%%%%%%%%%%%%%%%%%%%%%%%%%%%%%%%%%%%%%%%%%%%%%%%%%%%%%%%%%%%%%%%%%%%%%%%%%%%%%%%%%%%%%%%%%%%%%%%%%%%%%%%%%
\begin{figure}[h]
	\centering
	%\begin{tabular}{rl}
		\includegraphics[width=8cm, height=8cm]{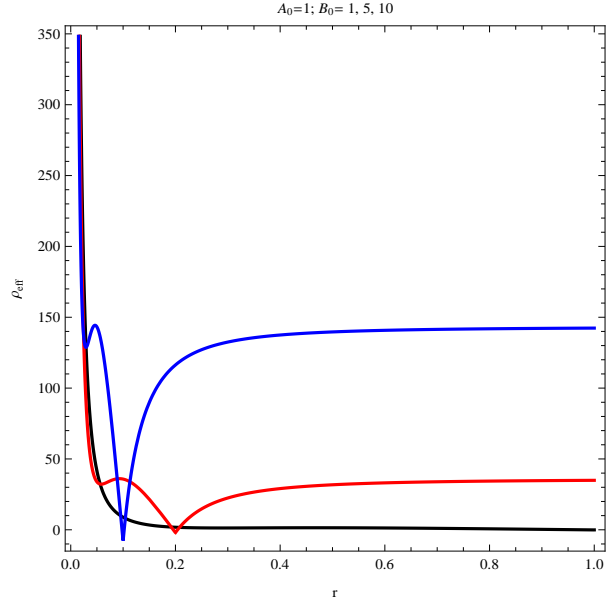}
		%&
		%\includegraphics[width=7cm, height=7cm]{somr01.eps}
			%\end{tabular}
	\caption{The curve indicates the variation of $\rho_{eff}$ versus radial coordinate, according to the gravitational model (\ref{model1}), with $C_1=1$, $A_0=1$, $B_0=1$ (black curve), $B_0=5$ (red curve) and $B_0=10$ (blue curve).}
	\label{fig1}
\end{figure}
%%%%%%%%%%%%%%%%%%%%%%%%%%%%%%%%%%%%%%%%%%%%%%%%%%%%%%%%%%%%%%%%%%%%%%%%%%%%%%%%%%%%%%%%%%%%%%%%%%%%%%%%%%%%
%%%%%%%%%%%%%%%%%%%%%%%%%%%%%%%%%%%%%%%%%%%%%%%%%%%%%%%%%%%%%%%%%%%%%%%%%%%%%%%%%%%%%%%%%%%%%%%%%%%%%%%%%%%%
%%%%%%%%%%%%%%%%%%%%%%%%%%%%%%%%%%%%%%%%%%%%%%%%%%%%%%%%%%%%%%%%%%%%%%%%%%%%%%%%%%%%%%%%%%%%%%%%%%%%%%%%%%%
%%%%%%%%%%%%%%%%%%%%%%%%%%%%%%%%%%%%%%%%%%%%%%%%%%%%%%%%%%%%%%%%%%%%%%%%%%%%%%%%%%%%%%%%%%%%%%%%%%%%%%%%%%%
\begin{figure}[h]
	\centering
	\begin{tabular}{rl}
		\includegraphics[width=8cm, height=8cm]{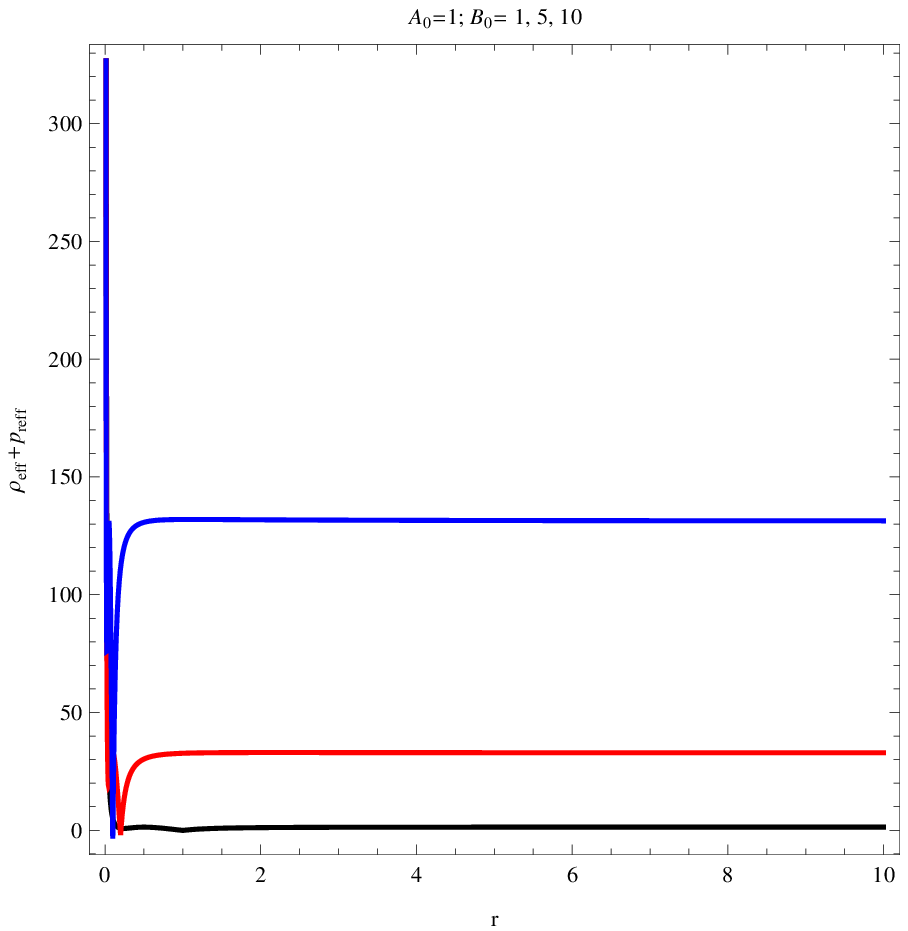}
		&
		\includegraphics[width=8cm, height=8cm]{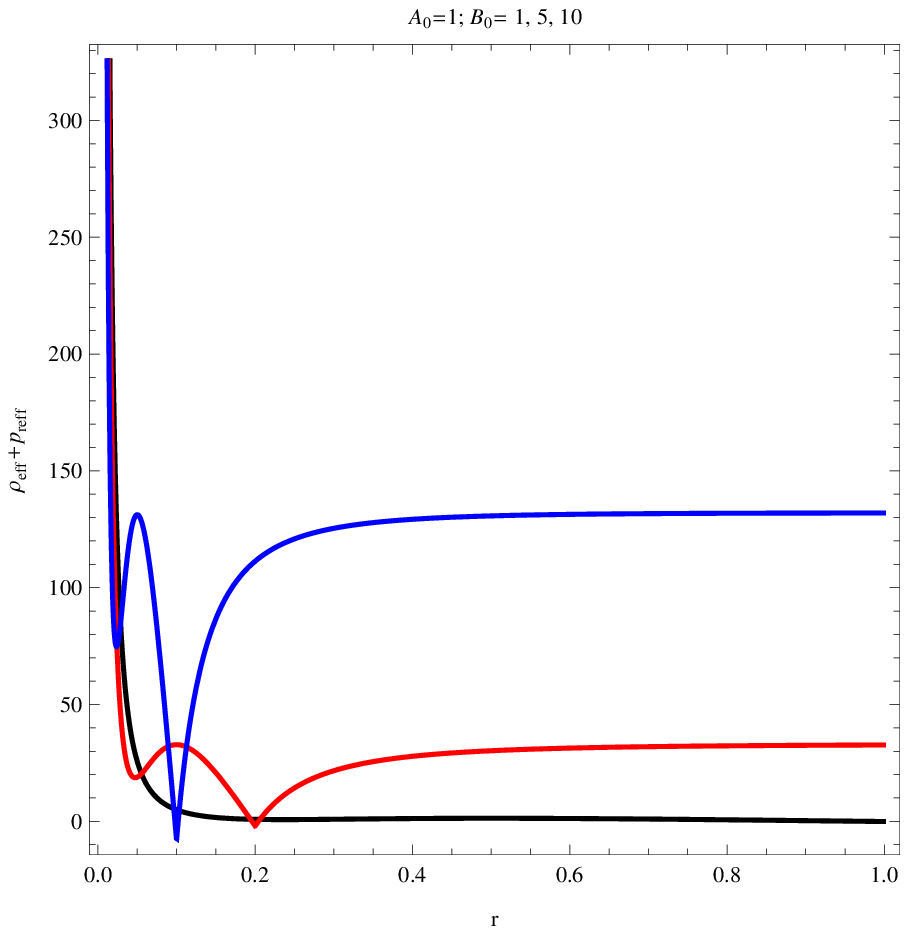}
			\end{tabular}
	\caption{Both panels indicate the variations of $\rho_{eff}+p_{r\,eff}$ versus radial coordinate, according to the gravitational model (\ref{model1}), with $C_1=1$, $A_0=1$, $B_0=1$ (black curve), $B_0=5$ (red curve) and $B_0=10$ (blue curve).}
	\label{fig2}
\end{figure}

\begin{figure}[h]
	\centering
	\begin{tabular}{rl}
		\includegraphics[width=8cm, height=8cm]{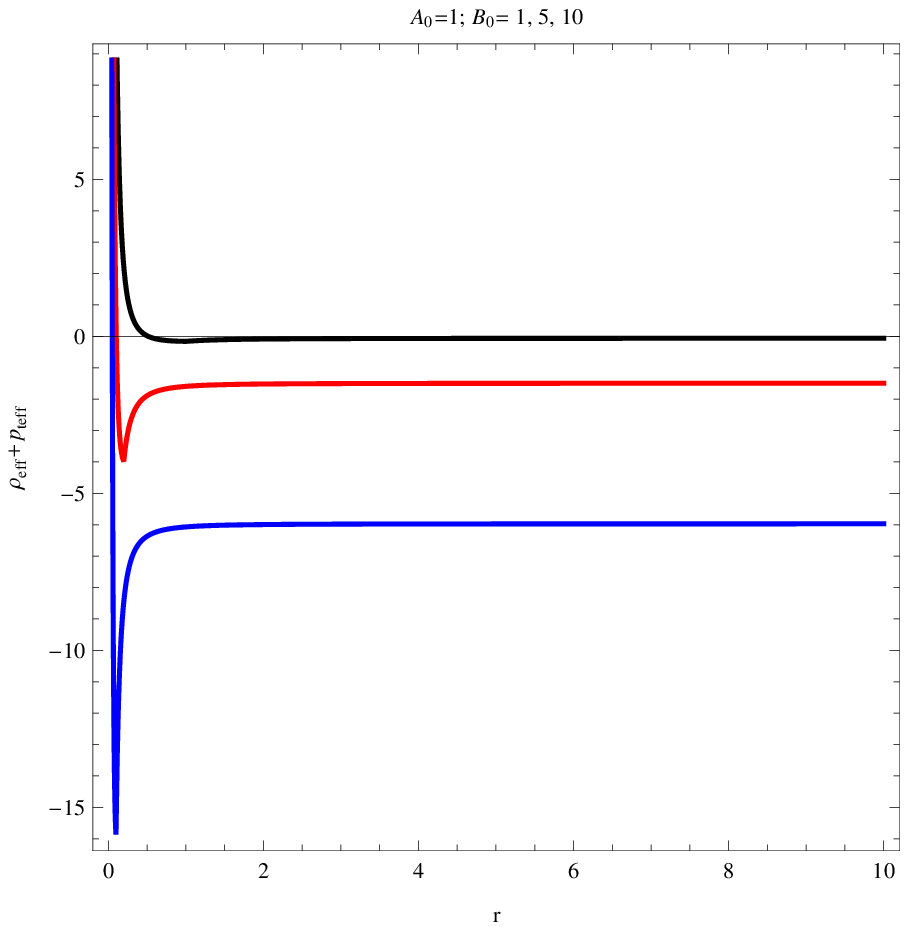}
		&
		\includegraphics[width=8cm, height=8cm]{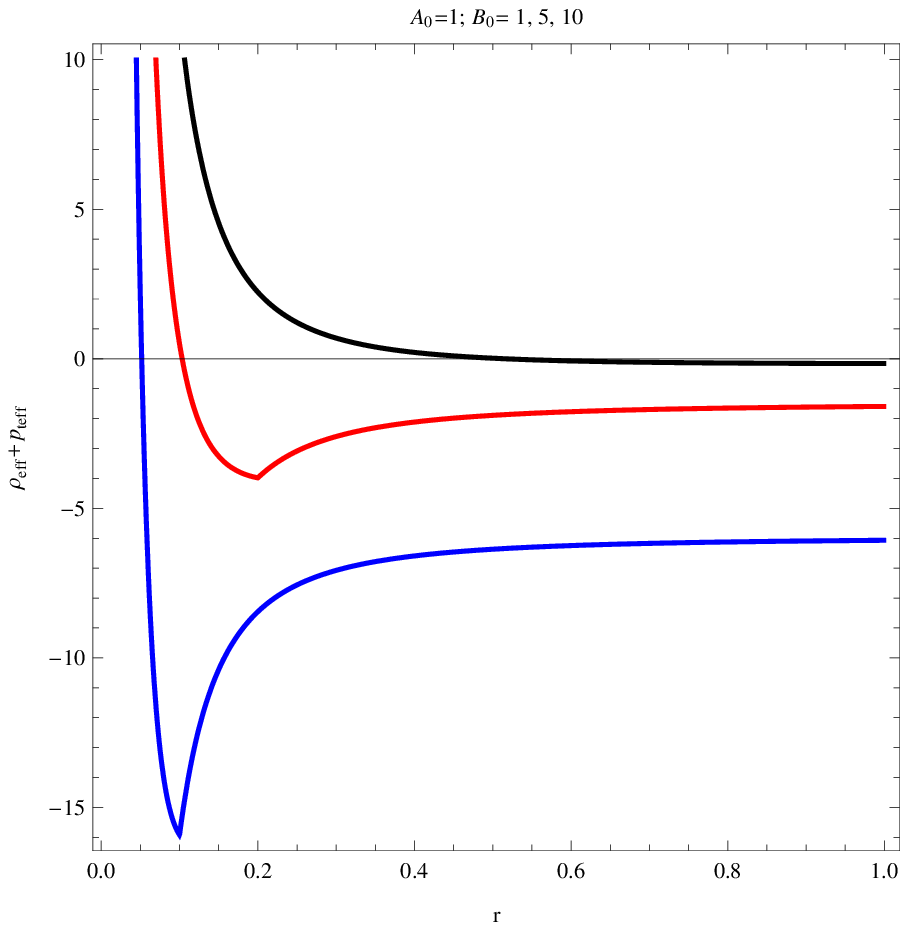}
			\end{tabular}
	\caption{ Both panels indicate the variations of $\rho_{eff}+p_{t\,eff}$ versus radial coordinate, according to the gravitational model (\ref{model1}), with $C_1=1$, $A_0=1$, $B_0=1$ (black curve), $B_0=5$ (red curve) and $B_0=10$ (blue curve). }
	\label{fig3}
\end{figure}

%\begin{figure}[h]
%	\centering
	%\begin{tabular}{rl}
	%	\includegraphics[width=7cm, height=7cm]{deffr010.eps}
	%	&
	%	\includegraphics[width=7cm, height=7cm]{deffr01.eps}
		%	\end{tabular}
	%\caption{ Both panels indicate the variations of $\rho_{eff}-p_{r\,eff}$ versus radial coordinate, according to the gravitational model (\ref{model1}), with $C_1=1$, $A_0=1$, $B_0=1$ (black curve), $B_0=5$ (red curve) and $B_0=10$ (blue curve). }
	%\label{fig3}
%\end{figure}

%\begin{figure}[h]
%	\centering
%	\begin{tabular}{rl}
	%	\includegraphics[width=7cm, height=7cm]{deffp010.eps}
	%	&
	%	\includegraphics[width=7cm, height=7cm]{deffp01.eps}
		%	\end{tabular}
	%\caption{ Both panels indicate the variations of $\rho_{eff}-p_{t\,eff}$ versus radial coordinate, according to the gravitational model (\ref{model1}), with $C_1=1$, $A_0=1$, $B_0=1$ (black curve), $B_0=5$ (red curve) and $B_0=10$ (blue curve). }
	%\label{fig4}
%\end{figure}
%%%%%%%%%%%%%%%%%%%%%%%%%%%%%%%%%%%%%%%%%%%%%%%%%%%%%%%%%%%%%%%%%%%%%%%%%%%%%%%%%%%%%%%%%%%%%%%%%%%%%%%%%%%%%%%%%%%%%%%%%%%%
%%%%%%%%%%%%%%%%%%%%%%%%%%%%%%%%%%%%%%%%%%%%%%%%%%%%%%%%%%%%%%%%%%%%%%%%%%%%%%%%%%%%%%%%%%%%%%%%%%%%%%%%%%%%%%%%%%%%%%%%%%%%%%%
%%%%%%%%%%%%%%%%%%%%%%%%%%%%%%%%%%%%%%%%%%%%%%%%%%%%%%%%%%%%%%%%%%%%%%%%%%%%%%%%%%%%%%%%%%%%%%%%%%%%%%%%%%%%%%%%%%%%%%%%%%%%%%

\begin{figure}[h]
	\centering
	%\begin{tabular}{rl}
		\includegraphics[width=8cm, height=8cm]{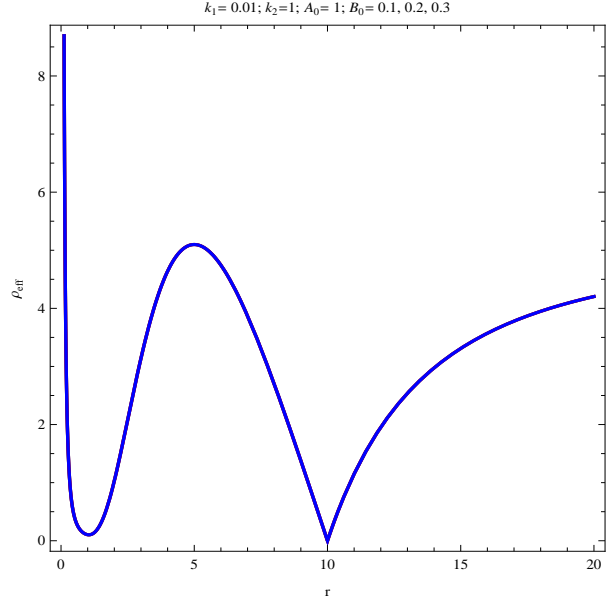}
		%&
		%\includegraphics[width=7cm, height=7cm]{som2r01.eps}
			%\end{tabular}
	\caption{ The curve indicates the variations of $\rho_{eff}$ versus radial coordinate, according to the gravitational model (\ref{gsol1}), with $C_1=1$, $k_1=0.01$, $k_2=1$, $A_0=1$, $B_0=0.1$ (black curve), $B_0=0.2$ (red curve) and $B_0=0.3$ (blue curve). }
	\label{fig4}
\end{figure}

%%%%%%%%%%%%%%%%%%%%%%%%%%%%%%%%%%%%%%%%%%%%%%%%%%%%%%%%%%%%%%%%%%%%%%%%%%%%%%%%%%%%%%%%%%%%%%%%%%%%%%%%%%%%
%%%%%%%%%%%%%%%%%%%%%%%%%%%%%%%%%%%%%%%%%%%%%%%%%%%%%%%%%%%%%%%%%%%%%%%%%%%%%%%%%%%%%%%%%%%%%%%%%%%%%%%%%%%%
%%%%%%%%%%%%%%%%%%%%%%%%%%%%%%%%%%%%%%%%%%%%%%%%%%%%%%%%%%%%%%%%%%%%%%%%%%%%%%%%%%%%%%%%%%%%%%%%%%%%%%%%%%%
%%%%%%%%%%%%%%%%%%%%%%%%%%%%%%%%%%%%%%%%%%%%%%%%%%%%%%%%%%%%%%%%%%%%%%%%%%%%%%%%%%%%%%%%%%%%%%%%%%%%%%%%%%%
\begin{figure}[h]
	\centering
	\begin{tabular}{rl}
		\includegraphics[width=8cm, height=8cm]{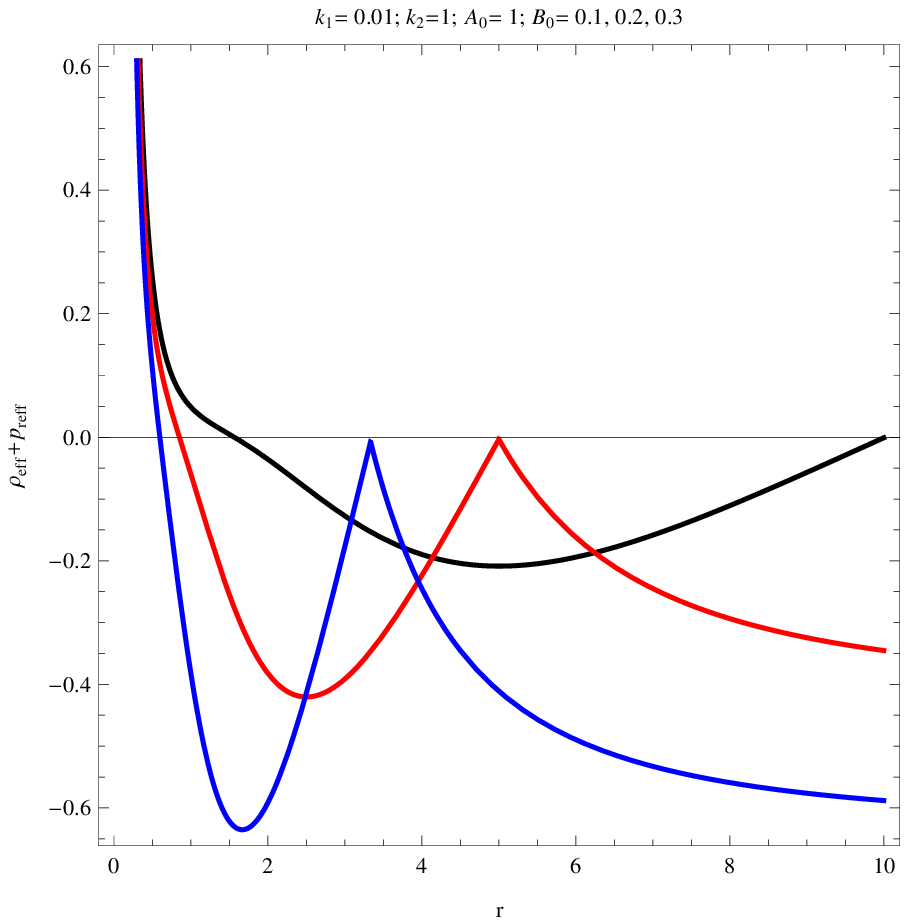}
		&
		\includegraphics[width=8cm, height=8cm]{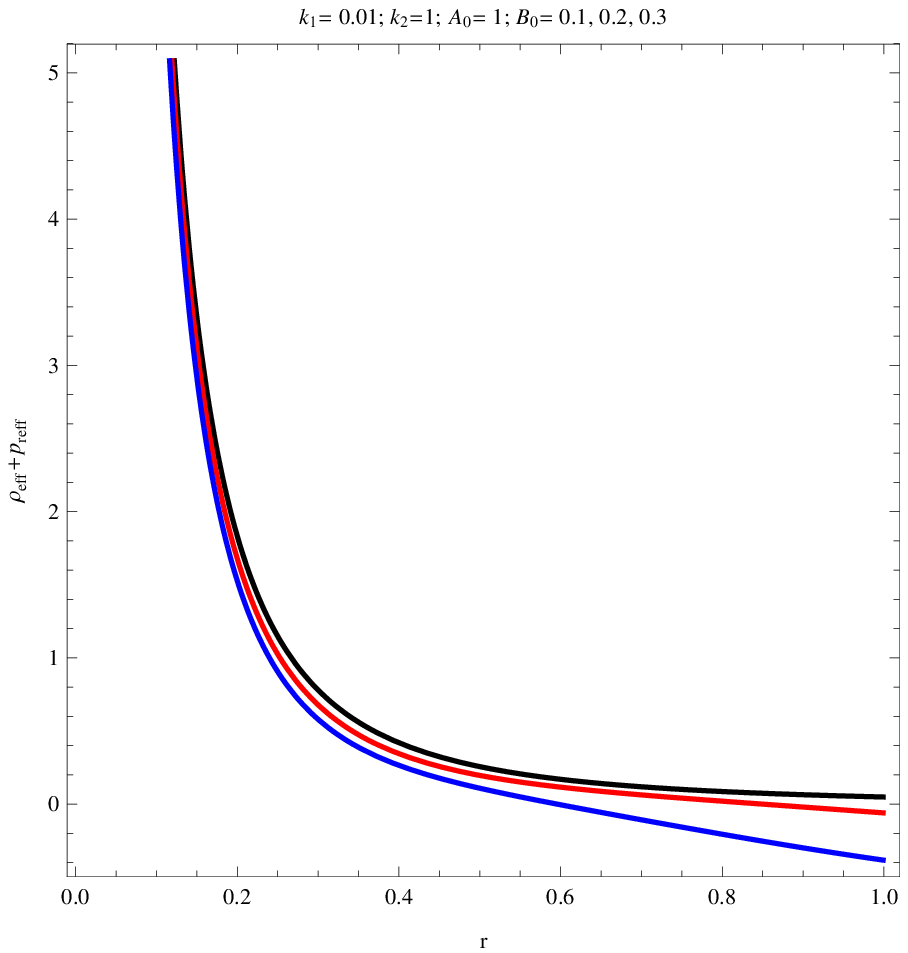}
			\end{tabular}
	\caption{ The left and right panels indicate the variations of $\rho_{eff}+p_{r\,eff}$ versus radial coordinate, according to the gravitational model (\ref{gsol1}), with $C_1=1$, $k_1=0.01$, $k_2=1$, $A_0=1$, $B_0=0.1$ (black curve), $B_0=0.2$ (red curve) and $B_0=0.3$ (blue curve). }
	\label{fig5}
\end{figure}

%%%%%%%%%%%%%%%%%%%%%%%%%%%%%%%%%%%%%%%%%%%%%%%%%%%%%%%%%%%%%%%%%%%%%%%%%%%%%%%%%%%%%%%%%%%%%%%%%%%%%%%%%%%%
%%%%%%%%%%%%%%%%%%%%%%%%%%%%%%%%%%%%%%%%%%%%%%%%%%%%%%%%%%%%%%%%%%%%%%%%%%%%%%%%%%%%%%%%%%%%%%%%%%%%%%%%%%%%
%%%%%%%%%%%%%%%%%%%%%%%%%%%%%%%%%%%%%%%%%%%%%%%%%%%%%%%%%%%%%%%%%%%%%%%%%%%%%%%%%%%%%%%%%%%%%%%%%%%%%%%%%%%
%%%%%%%%%%%%%%%%%%%%%%%%%%%%%%%%%%%%%%%%%%%%%%%%%%%%%%%%%%%%%%%%%%%%%%%%%%%%%%%%%%%%%%%%%%%%%%%%%%%%%%%%%%%

\begin{figure}[h]
	\centering
	\begin{tabular}{rl}
		\includegraphics[width=8cm, height=8cm]{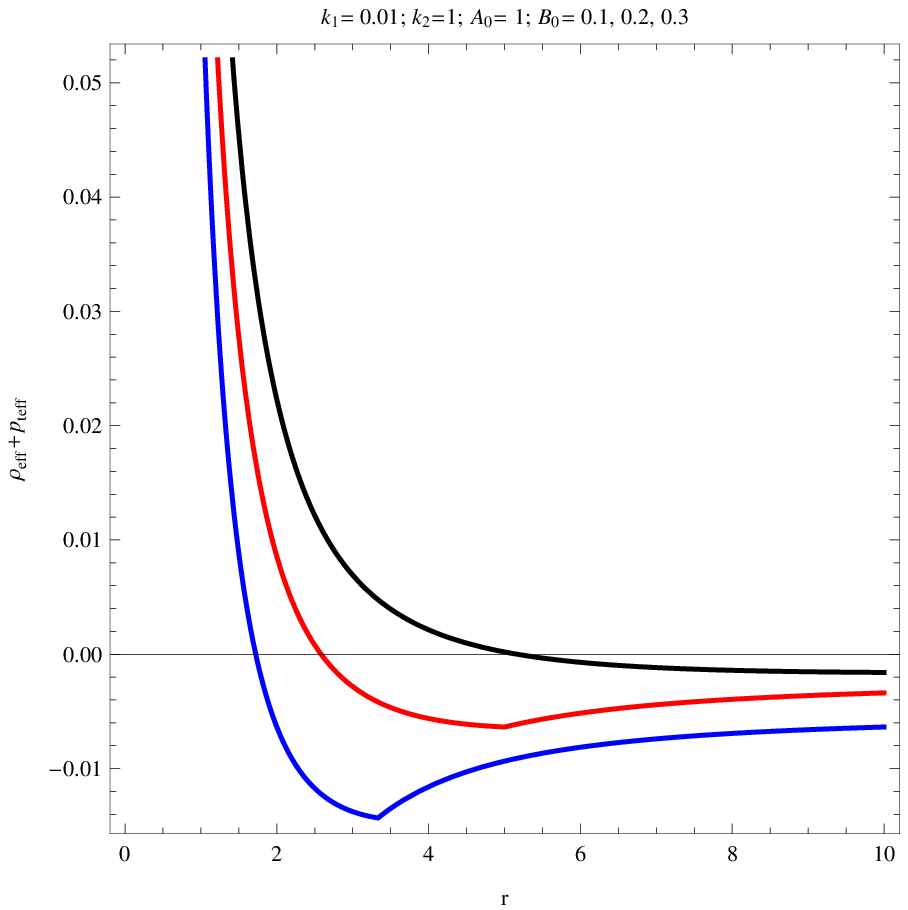}
		&
		\includegraphics[width=8cm, height=8cm]{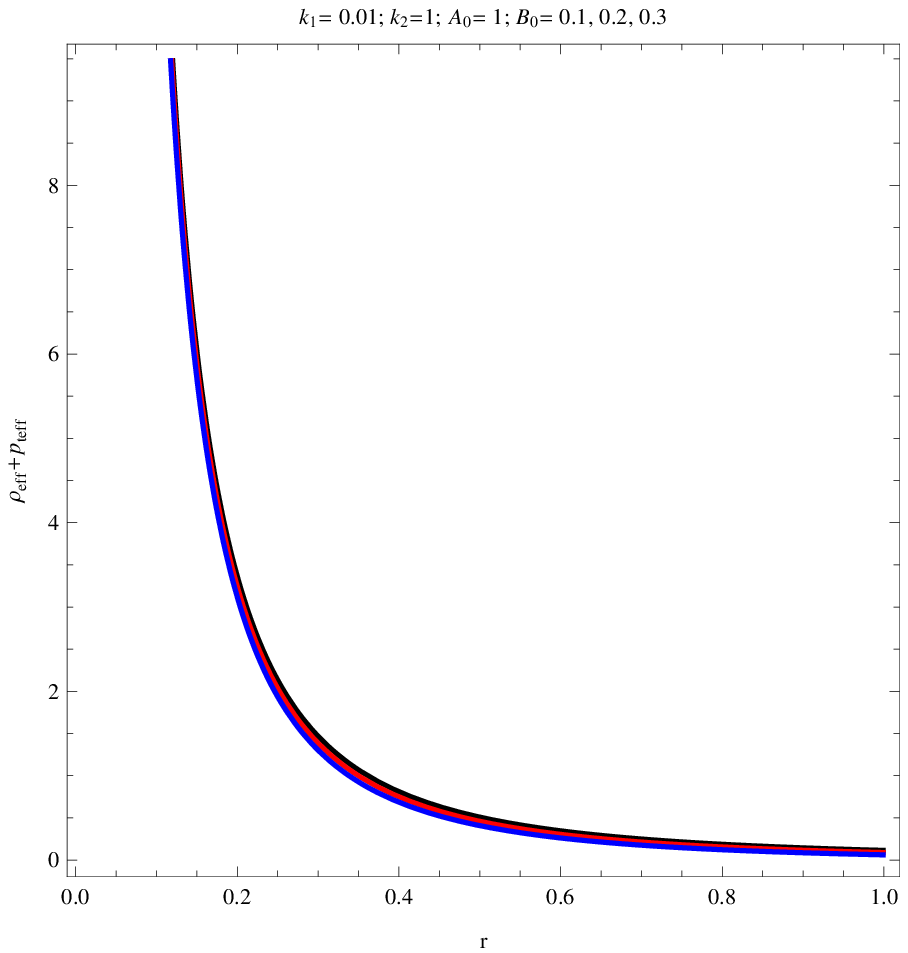}
			\end{tabular}
	\caption{ The left and right panels indicate the variations of $\rho_{eff}+p_{t\,eff}$ versus radial coordinate, according to the gravitational model (\ref{gsol1}), with $C_1=1$, $k_1=0.01$, $k_2=1$, $A_0=1$, $B_0=0.1$ (black curve), $B_0=0.2$ (red curve) and $B_0=0.3$ (blue curve).}
	\label{fig6}
\end{figure}

%\begin{figure}[h]
%	\centering
%	\begin{tabular}{rl}
%		\includegraphics[width=7cm, height=7cm]{deff2r10.eps}
%		&
%		\includegraphics[width=7cm, height=7cm]{deff2r01.eps}
%			\end{tabular}
%	\caption{ The left and right panels indicate the variations of $\rho_{eff}-p_{r\,eff}$ versus radial coordinate, according to the gravitational model (\ref{gsol1}), with $C_1=1$, $k_1=0.01$, $k_2=1$, $A_0=1$, $B_0=0.1$ (black curve), $B_0=0.2$ (red curve) and $B_0=0.3$ (blue curve). }
	%\label{fig7}
%\end{figure}

%\begin{figure}[h]
%	\centering
%	\begin{tabular}{rl}
	%	\includegraphics[width=7cm, height=7cm]{deff2p10.eps}
		%&
		%\includegraphics[width=7cm, height=7cm]{deff2p01.eps}
			%\end{tabular}
	%\caption{ The left and right panels indicate the variations of $\rho_{eff}-p_{t\,eff}$ versus radial coordinate, according to the gravitational model (\ref{gsol1}), with $C_1=1$, $k_1=0.01$, $k_2=1$, $A_0=1$, $B_0=0.1$ (black curve), $B_0=0.2$ (red curve) and $B_0=0.3$ (blue curve). }
	%\label{fig8}
%\end{figure}
With the reconstructed model (\ref{model1}) we perform the analyze about the energy conditions, more precisely the NEC.
The model depends on the parameters $C_1$, $A_0$ and $B_0$, is used to graphically study the evolution of $\rho_{eff}+p_{eff}$ and    $\rho_{eff}-p_{eff}$ for both the radial and tangential pressures. We fix $C_1=1$, $A_0=1$ and $B_0=1, 5, 10$. From Fig. \ref{fig1} and Fig. \ref{fig2}, it appears that $\rho_{eff}\geq 0$ and $\rho_{eff}+p_{r\,eff}\geq 0$, showing that the existence of the static traversable wormhole is possible even so null energy condition is satisfied is the radial direction. The Fig. \ref{fig3} indicates the evolution of $\rho_{eff}+p_{t\,eff}$ according to what it appears that in the tangential direction, the NEC should be violated, despite the existence of static traversable wormhole.  It can be concluded that the NEC is not determinant for the existence of static traversable wormhole. \par
According to the model (\ref{gsol1}), the NEC is also analyzed, the dependence is now characterized by the parameters fixed as $k_1=0.01$, $k_2=1$, $A_0=1$ and $B_0=0.1, 0.2, 0.3$. Hence, from Fig. \ref{fig4}, one always gets the condition $\rho_{eff}\geq 0$, whereas  for both the radial and tangential directions, the NEC should be violated despite the existence of the static traversable wormhole, as shown by Fig. \ref{fig5} and  Fig. \ref{fig6}. \par 
According to the results obtained in \cite{chakra}, it is concluded that, in $f(R)$ theory of gravity, energy conditions related to the matter threading the wormhole are in general found to obey the null energy conditions (NEC) in regions around the throat, while in the limit $f(R) = R$; NEC can be violated at large in regions around the throat. On the other hand, it is shown in \cite{sharif} that wormhole solutions can be obtained in the absence of exotic matter for some particular regions of spacetime; also in \cite{zia}, it has been analyzed with the graphical evolution that the null energy and weak energy conditions for the effective energy-momentum tensor are 
usually violated for the ordinary matter content, while some small feasible regions for the existence of wormhole solutions have been found where the energy conditions are not violated. Thus it comes from the various results that in some cases, the NEC can be (or not) violated for the existence of the wormhole. According to $f(T)$ theory of gravity performed in the present paper, the violation of the NEC is not a determinant condition for the existence of static traversable wormhole.

\newpage

\section{Conclusion}\label{sec5}
  This paper is devoted to the study of dynamical traversable wormhole solutions in the framework of $f(T)$ gravity, where $T$ denotes the torsion scalar. We then consider the metric (\ref{le}), characteristic of dynamic and traversable wormhole.  Within diagonal tetrads treatment, constraints reduce the algebraic $f(T)$ to the standard teleparallel gravitational action term, where the results are quite trivial. In order to get interesting results we focus on non-diagonal tetrads, and according to specific shape functions, $f(T)$ models are reconstructed.   The first important and interesting remark of obtaining non-teleparallel term is impossibility of getting time dependent scale factor, that is $A(t)$ has to be constant, constraining the possible traversable wormholes to be static.\par 
	Our study is also extended to the analysis of the energy conditions around the reconstructed $f(T)$ models. With the model (\ref{model1}) and for some values of the input parameters the NEC is always satisfied in the radial direction, while in the tangential direction the NEC should be violated, even so the existence the static traversable wormhole. However, with the model (\ref{gsol1}), it appears from the Fig. \ref{fig5} and Fig. \ref{fig6} that the NEC should be violated the static traversable wormhole exists. Thus, we conclude that the violation of the null energy condition condition is not determinant for the existence of static traversable wormhole.

%\newpage

%\section*{Acknowledgments}

\end{document}